\documentclass[aip,amsmath,amssymb,reprint,floatfix]{revtex4-1}

\usepackage{graphicx}
\begin{document}


\title{On the Validity and Applicability of Models of Negative Capacitance and Implications for MOS Applications} 



\author{J. A. Kittl}
\author{B. Obradovic}
\author{D. Reddy}
\author{T. Rakshit}
\author{R. M. Hatcher}
\author{M. S. Rodder}
\affiliation{%
Advanced Logic Lab, Samsung Semiconductor, Inc., Austin, TX 78754, USA
}%

\date{\today}

\begin{abstract}
\noindent The observation of room temperature sub-60 mV/dec subthreshold slope (SS) in MOSFETs with ferroelectric (FE) layers in the gate stacks or in series with the gate has attracted much attention. Recently, we modeled this effect in the framework of a FE polarization switching model. However, there is a large amount of literature attributing this effect to a stabilization of quasi-static (QS) negative capacitance (NC) in the FE. The technological implications of a stabilized non-switching (NS) QSNC model vs a FE switching model are vastly different; the latter precluding applications to sub-60 mV/dec SS scaled CMOS due to speed limitations and power dissipated in switching. In this letter, we provide a thorough analysis assessing the foundations of models of QSNC, identifying which specific assumptions (ansatz) may be unlikely or unphysical, and analyzing their applicability. We show that it is not reasonable to expect QSNC for two separate capacitors connected in series (with a metal plate between dielectric (DE) and FE layers). We propose a model clarifying under which conditions a QS ``apparent NC'' for a FE layer in a FE-DE bi-layer stack may be observed, quantifying the requirements of strong interface polarization coupling in addition to capacitance matching. In this regime, our model suggests the FE layer does not behave as a NC layer, simply, the coupling leads to both the DE and FE behaving as high-k DE with similar permittivities. This may be useful for scaled equivalent oxide thickness (EOT) devices but does not lead to sub-60 mV/dec SS.
\end{abstract}

\pacs{}

\maketitle 

\begin{figure}[b]
	\centering
		\includegraphics[width=0.35\textwidth]{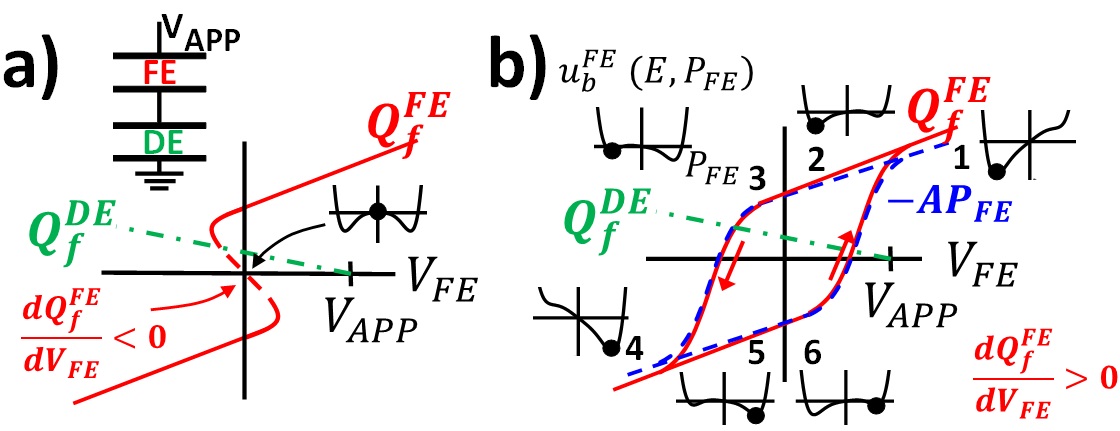}
	\caption{Free charge vs voltage of a FE capacitor according to (a) stabilized QS negative capacitance models and (b) conventional hysteretic behavior.}
	\label{fig:fig_1}
\end{figure}

\begin{figure}
	\centering
		\includegraphics[width=0.27\textwidth]{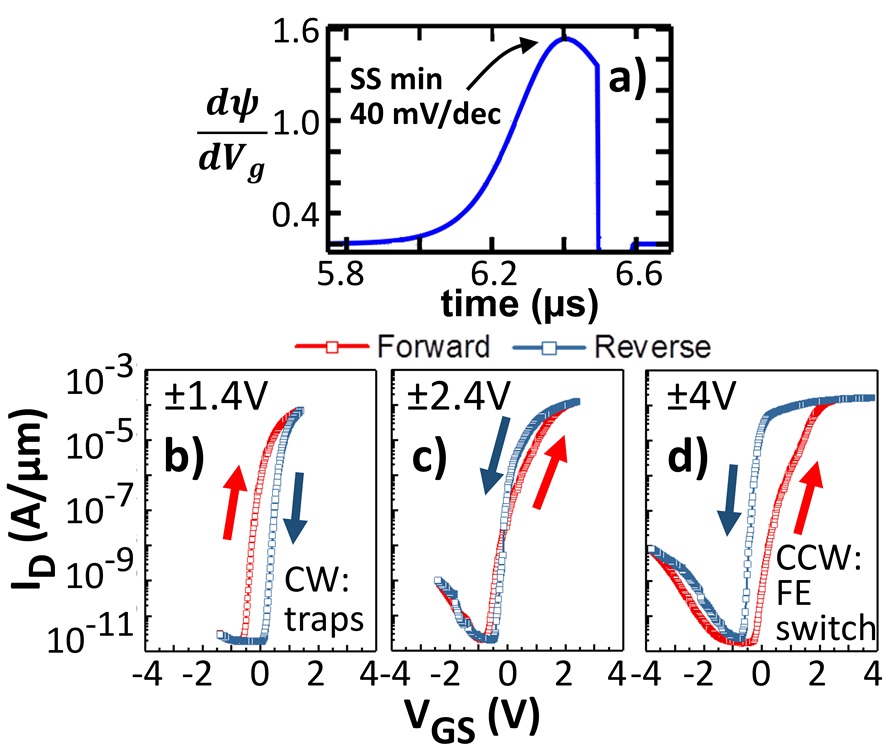}
	\caption{For MOS devices with a FE layer in the gate stack a) a model incorporating a delay in FE polarization switching predicts sub-60 mV/dec SS and b, c and d) compensation of CW trapping-detrapping hysteresis and CCW FE switching hysteresis can result in low hysteresis.}
	\label{fig:fig_2}
\end{figure}

The occurrence of stabilized quasi-static (QS) negative capacitance (NC) in systems that include ferroelectric (FE) layers has recently been postulated \cite{salahuddin_use_2008,khan_ferroelectric_2011,salvatore_experimental_2012,jain_stability_2014} and been the subject of many studies \cite{salahuddin_use_2008,khan_ferroelectric_2011,salvatore_experimental_2012,jain_stability_2014,jo_experimental_2015,li_sub-60mv-swing_2015,nourbakhsh_subthreshold_2017,krivokapic_14nm_2017,kwon_improved_2018,majumdar_revisiting_2016,khan_work_2017,rollo_energy_2017,dong_simple_2017,agarwal_designing_2018,chang_thermodynamic_2017}. The observation of room temperature sub-60 mV/dec sub-threshold slope (SS) for MOSFET devices that included FE layers in the gate stack or in series with the gate \cite{sharma_impact_2017,sharma_time-resolved_2018,jo_experimental_2015,li_sub-60mv-swing_2015,nourbakhsh_subthreshold_2017} has triggered significant interest due to potential applications in low power CMOS. Observation of sub-60 mV/dec SS \cite{jo_experimental_2015,li_sub-60mv-swing_2015,nourbakhsh_subthreshold_2017} or improvements of SS \cite{krivokapic_14nm_2017,kwon_improved_2018} have been regarded as supporting evidence for models of stabilized QSNC \cite{salahuddin_use_2008,khan_ferroelectric_2011,salvatore_experimental_2012,jain_stability_2014,jo_experimental_2015,krivokapic_14nm_2017,majumdar_revisiting_2016,khan_work_2017,rollo_energy_2017,dong_simple_2017,agarwal_designing_2018,chang_thermodynamic_2017}. In these models, the FE layer is proposed to traverse quasi-statically and reversibly a region of negative capacitance (dotted line in Fig. \ref{fig:fig_1}a) when in series with a dielectric (DE) layer or capacitor, and under a ``capacitance matching'' condition that stabilizes this path with no FE switching (NS), instead of a conventional FE hysteretic path based on FE switching (solid line in Fig. \ref{fig:fig_1}b)  \cite{salahuddin_use_2008,khan_ferroelectric_2011,salvatore_experimental_2012,jain_stability_2014,jo_experimental_2015,krivokapic_14nm_2017,majumdar_revisiting_2016,khan_work_2017,rollo_energy_2017,dong_simple_2017,agarwal_designing_2018,chang_thermodynamic_2017}. Recently, we have proposed an alternative explanation for the experimental observation of sub-60 mV/dec SS in devices containing FE layers \cite{Obradovic_Modeling_2018} (Fig. \ref{fig:fig_2}), attributing the sub-60 mV/dec SS to effects resulting from FE polarization switching and transient dynamic NC to a switching delay. Switching kinetics for HfO$_{2}$-based FE (used extensively in device studies) was measured for scaled devices (80 nm width and 30 nm length) \cite{mulaosmanovic_switching_2017} finding that switching is quite slow, with time constants of $\sim$ ms-$\mu$s at voltages of 2-3 V for 10 nm films \cite{mulaosmanovic_switching_2017}. We attribute the small or apparent lack of hysteresis observed in some experiments (considered as supporting evidence of NS models) to canceling  between the counter-clockwise (CCW) FE  and the clockwise (CW) charge trapping-detrapping hysteresis (See examples in Fig.2, measured on samples fabricated following experimental details given in \cite{sharma_impact_2017, sharma_time-resolved_2018}). The implications of the stabilized NS QSNC model vs. those of FE switching-based models for sub-60 mV/dec SS MOS applications are significantly different. The former predicts no significant operation speed limitations and would result in power reduction. In contrast, FE switching-based models predict several drawbacks to sub-60 mV/dec SS CMOS operation, including limiting operating speeds to values consistent with FE switching (kHz-MHz range clock frequencies for HfO$_{2}$-based FEs), and additional power dissipated in switching. We note that devices (e.g. in ring oscillators) with FE in the gate stack can still be operated at higher frequencies, but the FE switching response will not follow (no sub-60 mV/dec SS), rather, they will show conventional (DE gate stack-like) behavior. 

In this letter, we provide a thorough analysis assessing the foundations of models of stabilized NS QSNC, identifying which specific assumptions (ansatz) may be unlikely or unphysical, clarifying the difficulties with these assumptions, and analyzing their applicability to specific systems. We also propose a model that can result in ``apparent'' NC on FE-DE bi-layers under a condition of strong interface polarization coupling in addition to capacitance matching; in this regime, the FE layer does not behave as a NC layer, rather, the coupling leads to both DE and FE layers behaving as high-k DE with similar permittivities. This may be useful for scaled EOT devices but does not lead to sub-60 mV/dec SS.

\begin{figure}[b]
	\centering
		\includegraphics[width=0.3\textwidth]{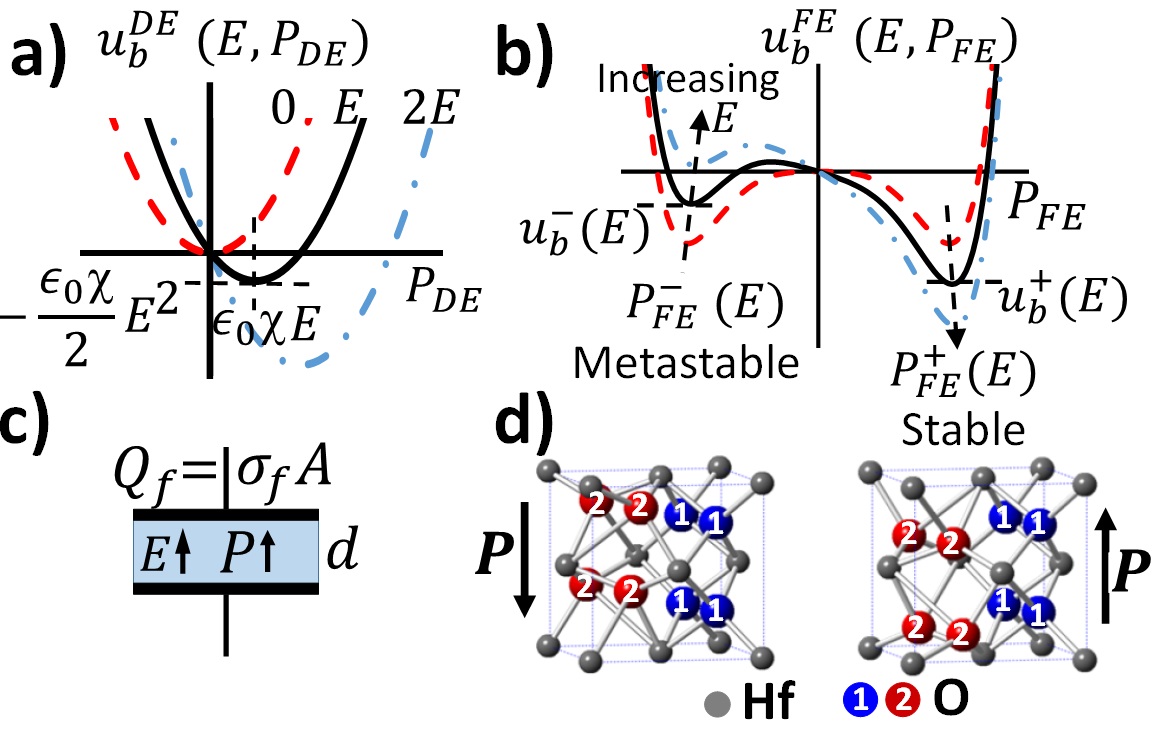}
	\caption{Schematics of internal (configuration) free energy of a) linear dielectric, and b) ferroelectric materials vs. polarization. c) Planar capacitor considered in the analysis, d) atomic configurations of FE HfO$_2$.}
	\label{fig:fig_3}
\end{figure}

We first consider the properties of linear dielectric (DE) and FE materials (Fig. 3). For simplicity, we assume fields and polarizations in the z-direction. We consider the free energy per unit volume of material, $u_b$, as a function of its polarization, $P$. The arguments made here are general, but in order to illustrate the issues with previous models of negative capacitance which use Landau's phenomenological mean field description, we follow this approach: 
\begin{equation}
	u_b(P)=\frac{{\alpha }}{2}P^2+\ \frac{{\beta }}{4}P^4+\ \frac{{\gamma }}{6}P^6-EP
	\label{eq:UFE}
\end{equation}
\noindent where ${\alpha}$, ${\beta}$ and ${\gamma}$ are material parameters and $E$ is the applied field. For linear DE ${\alpha }_{DE} > 0 $, ${\beta}_{DE}={\gamma}_{DE}=0$, and ${\alpha }_{DE} = {1}/{\chi \epsilon_{0}}$ with $\chi$: susceptibility and $\epsilon_{0}$: vacuum permittivity. Alternatively, the external field can be used in the formalism, and is typically used for FE with well-established parameters for typical FE, e.g.,\cite{qu_interfacial_1997} with ${\alpha }_{FE}<0$ and ${\beta }_{FE}$ or ${\gamma }_{FE} > 0 $. Both formalisms (using applied or external field) have been used in QSNC models. Conclusions of this paper are agnostic to the choice of formalism. Note that the energies and coefficients are different for the formalisms using applied vs external field, and that minimization is only valid when keeping the applied or external field constant respectively (by analogy to Helmholtz and Gibbs free energies not being the same, and minimization only being valid at constant volume or pressure respectively). A set of arguments for QSNC based on the formalism with applied field point out that a stabilization of the FE layer around $P=0$ (energy minimum) by itself (w/o having a ``stabilizing'' DE layer) could happen for ${-2}/{\epsilon }_0<{\alpha }_{FE}< 0$ \cite{hoffmann_stabilization_2018}. These arguments ignore the need to maintain the applied field over the FE constant in the minimization. In any case, this would be a material in which the spontaneous polarization would not withstand its own depolarization field, would behave as a dielectric at small external fields with positive capacitance, and not lead to QS sub-60 mV/dec SS. This is similar to unpoled FE, that may have a dielectric type behavior around $P=0$, and cannot lead to QSNC \cite{krowne_examination_2011}. Using the formalism with the external field (as common for poled FE) and ${\alpha }_{FE}<0$, the argument falls apart in any case. Poled FE always withstands their own depolarization field, and the state around $P=0$ is always unstable. Eq \eqref{eq:UFE} indicates the theoretical values of $u_{b}$ that the material would have if taking a specific polarization under a field. From these possible polarization values, only those corresponding to  $u_{b}$ minima are physically observed under QS conditions. For the DE, there is a single equilibrium polarization possible under an applied field: $P_{DE}(E)=\chi {\epsilon }_0E$, corresponding to the energy minimum $u^{DE}_b\left(E\right)=-{(\epsilon }_0\chi /2)\ E^2.$ For the FE, there are two minima, a global and a local minimum corresponding to stable and metastable equilibrium respectively (Fig. \ref{fig:fig_3}b). Microscopically, this typically corresponds to two non-centrosymmetric atomic configurations as illustrated in Fig. \ref{fig:fig_3}d for FE HfO$_{2}$ \cite{park_ferroelectricity_2015}, in which there are atomic displacements (either in  +z or -z direction) from the centrosymmetric configuration. The case of $P_{FE}=0$ (centrosymmetric configuration) is microscopically unstable, and consequently not a physically possible QS configuration. This leads to the hysteretic behavior of FE (Fig. \ref{fig:fig_1}b). It is also known that FE materials have domains within which the polarizations are aligned. A statistical analysis (e.g. following the Preisach model \cite{bartic_preisach_2001}) adequately describes the QS behavior of multi-domain FE. Macroscopically, $P_{FE}=0$ is achieved as an average over domains or during polarization switching. 

We now consider the QS behavior of a planar capacitor of area $A$ with a single material (DE or FE) of thickness $d$ between its metal plates and a free charge density ${\sigma }_f=Q_f/A$ in its plates (Fig. \ref{fig:fig_3}c). For a DE, $\sigma_{f}$ determines uniquely the polarization state $P_{DE}\left({\ \sigma }_{f\ }\right)=~-({\epsilon }_0/\epsilon){\chi}{\sigma}_{f}$, where $\epsilon={\epsilon }_{0}{(1+ \chi)}$ is the DE permittivity. For a given free charge in the plates of the capacitor, the material only exhibits the polarization which minimized its free energy (only this polarization is physically realized). Similarly, for FE capacitors, at a given $Q_f$ in the plates, the material adopts a local or global minimum in free energy: only the polarizations of metastable or stable minima are observed, as described by the hysteretic FE behavior.  The history of free charges, $Q_f(t)$, uniquely determines the polarization state of the FE capacitor (including poling, cycling, etc.). 

Simply stated: under QS conditions, we are considering one internal degree of freedom ($P$) for a given single material in a capacitor with metal plates. There is also only one external degree of freedom (that can be controlled externally) in our example: either the applied field (voltage) history or the free charge history can be controlled as imposed parameters. Under either one of these external constraints, free energy minimization determines univocally the polarization history ($P(t)$), and the history of the remaining parameter is also determined univocally through $Q_f=A(\epsilon_{0}V/d-P)$. For a FE capacitor, this results in a conventional hysteretic behavior path as response to any arbitrary applied $V(t)$ history or to any arbitrary $Q_{f}(t)$ history. There is simply no room (no degrees of freedom left) for an alternative (non-hysteretic) path for any given $V(t)$ or $Q_{f}(t)$ history.

\begin{figure}[ht]
	\centering
		\includegraphics[width=0.35\textwidth]{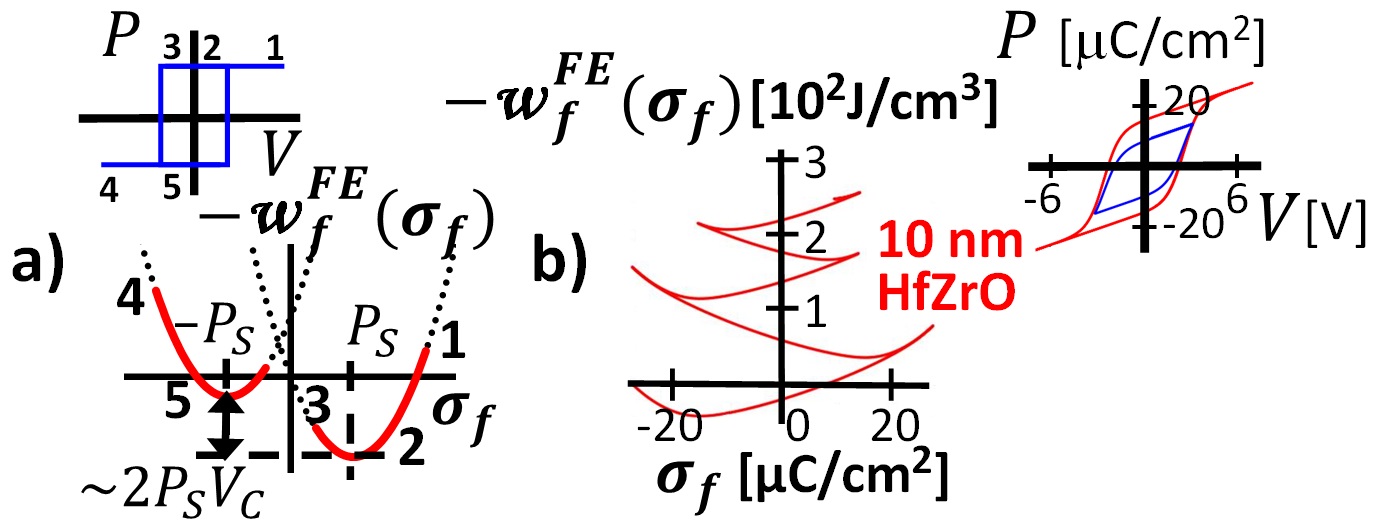}
	\caption{Work ($w_{f}^{FE}$) performed to form a free charge configuration in a FE capacitor (normalized to volume of FE): a) Schematic for ideal ``hysteron'' and b) illustrating the case of FE (Hf-Zr)O$_{2}$. Second derivatives of ${-w}^{FE}_f\left({\ \sigma }_{f\ }\right)$ are always positive, indicating only positive capacitances are observed.}
	\label{fig:fig_4}
\end{figure}

We now consider the work performed to form a configuration of free charges in the capacitor, $W_f$:
\begin{equation} 
-Ad\ w_f\left({\ \sigma }_f\right)=-W_f\left(Q_f\right)=\int{V(Q_f)~dQ_f} 
\label{eq:workF} 
\end{equation} 
where $V$ is the applied voltage. The capacitance is: $C=\ -({A}/{d}) ({{\partial }^2w_f}/{\partial {\sigma }^2_f})^{-1}$ . For a DE capacitor:
\begin{equation}
u^{DE}_f\left({\ \sigma }_{f\ }\right)\equiv -w^{DE}_f\left({\ \sigma }_{f\ }\right)={{\sigma }^2_f}/{2\epsilon }
\label{eq:UDEsigma}
\end{equation} 
where $u^{DE}_f\left({\ \sigma }_{f\ }\right)$ is the energy of the free charge configuration in the plates of the DE capacitor.
\noindent For a FE capacitor, we consider first an ideal ``hysteron'', i.e. $P-V$ relation with abrupt switching at  $\pm V_C$  (coercive voltage) between two values of saturation polarization $\pm P_S$. Fig. \ref{fig:fig_4}a shows that  ${-w}^{FE}_f\left({\ \sigma }_{f\ }\right)$ for the ideal ``hysteron'' is composed of parabolic branches  centered around $P_S$ and${-P}_S$. Each time the polarization switches, ${\ \sigma }_{f\ }$ jumps (right or left depending on direction of switching) by $2P_S$, and the energy jumps (``climbs'') by $2P_SV_C$, due to the energy dissipated in switching (total dissipation of $4P_SV_C$ for the full hysteresis cycle). The second derivative of ${-w}^{FE}_f\left({\ \sigma }_{f\ }\right)$ is always positive: FE capacitors cycled under QS conditions always have positive capacitance. Fig. \ref{fig:fig_4}b illustrates a more realistic case of cycling a (Hf-Zr)O$_{2}$ FE capacitor, with minor and major loops during cycling. The main effects described for the ``ideal'' hysteron are observed: the second derivative of ${-w}^{FE}_f\left({\ \sigma }_{f\ }\right)$ is always positive (always positive capacitance), and the ``climbing'' through cycling due to hysteretic losses in switching. The curve is now smooth since the ideal hysteron  is replaced by a realistic and continuous \textit{P-V } relation. Regions of negative second derivative of $u_{b}(P)$ do not imply QSNC (negative second derivative of $u_{f}(\sigma_{f})$. To emphasize this, we consider the QS capacitance $C={{\Delta}Q_{f}}/{{\Delta}}V$ (small  ${\Delta}V=V_{2}-V_{1}$, where ${\Delta}Q=Q_{f2}-Q_{f1}$, states 1 and 2 are equilibrium states (stable or metastable) and the system evolves from 1 to 2). For a FE capacitor, comparing states 1 and 2, we always have ${{\Delta}P_{FE}}/{{\Delta}E}\geq 0$ (where ${\Delta}P_{FE}=P_{FE2}-P_{FE1}$, and ${\Delta}E=E_{2}-E_{1}$) both for the dielectric component (small displacements around eq. positions) and for the FE switching component. Since $C=(A/d)[\epsilon_{0}+({\Delta}P_{FE}/{\Delta}E)]$, the QS capacitance of a FE capacitor is always positive. In the case FE switching events take place, the barrier to FE switching, determined by the shape of $u_{b}(P)$ close to $P=0$, does not play any role in determining the changes in $P_{FE}$, $Q_{f}$ or $V$ between states 1 and 2. Transition barriers play a role in kinetics (dynamics), not on QS considerations. 

\begin{figure}[ht]
	\centering
		\includegraphics[width=0.3\textwidth]{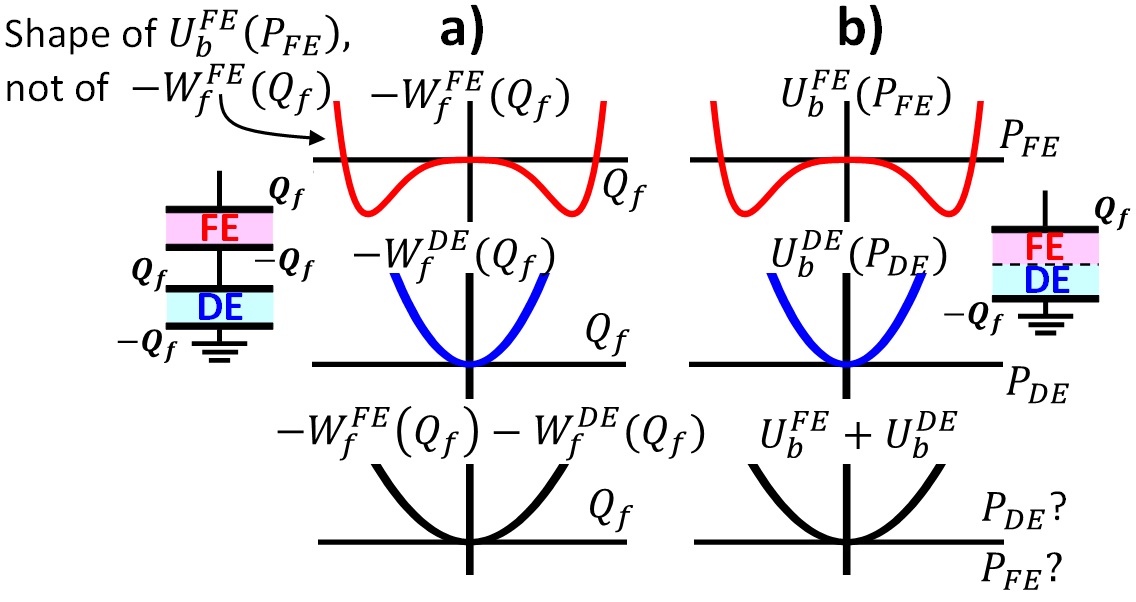}
	\caption{a) Negative capacitance ``ansatz 1'', for DE and FE capacitors connected in series, assumes an incorrect shape for $-W_{f}^{FE}$ vs $Q_f$ with a negative second derivative section around $Q_f=0$. b) Negative capacitance ``ansatz 2'', for a DE-FE bilayer capacitor, in order to be able to add the free energies of the FE and DE plotted vs. the same polarization axis, would require the polarization of FE and DE layers to be the same (not observed in general).}
	\label{fig:fig_5}
\end{figure}

We are now ready to address the issues with proposed models of NS QSNC. The first case we consider is that of separate FE and DE capacitors connected in series (each one with metal plates) \cite{khan_ferroelectric_2011,salvatore_experimental_2012,jain_stability_2014,jo_experimental_2015,majumdar_revisiting_2016,khan_work_2017,rollo_energy_2017,agarwal_designing_2018,chang_thermodynamic_2017}. With the explanations presented above, it should suffice to conclude that it is not possible to achieve QSNC on the FE capacitor: the electric field in the metal plates of the FE capacitor under QS condition is 0, the FE layer is unaware of what is connected in series, other than through  the history of charges${\ Q}_f(t)$ at the internal interface in the metal plates of the FE capacitor, which completely and uniquely determines its behavior. The same hysteretic behavior as described above and observed in all FE capacitors for any arbitrary${\ Q}_f(t)$ is expected. Still, multiple publications describe and model QS stabilization of NC in such systems \cite{khan_ferroelectric_2011,salvatore_experimental_2012,jain_stability_2014,jo_experimental_2015,majumdar_revisiting_2016,khan_work_2017,rollo_energy_2017,agarwal_designing_2018,chang_thermodynamic_2017}. We here analyze the ansatz that leads to this (incorrect) conclusion (Fig. \ref{fig:fig_5}a). The ``QSNC ansatz 1'' considers the free energy of the free charge configurations ($-W_f\left({\ \sigma }_{f\ }\right)$) of two capacitors connected in series as a function of the free charge. The same free charge is present in both capacitors connected in series, so the free energies are plotted vs. the same axis  $Q_f$ (or ${\ \sigma }_{f\ }$). The total energy of the system is obtained as the sum of the energies of each capacitor. The problem, as illustrated in Fig. \ref{fig:fig_5}a, is that these models assume \cite{salahuddin_use_2008,jain_stability_2014,majumdar_revisiting_2016,khan_work_2017,rollo_energy_2017,dong_simple_2017,agarwal_designing_2018,chang_thermodynamic_2017} a form for ${-w}^{FE}_f\left({\ \sigma }_{f\ }\right)$ , e.g. as given by:
\begin{equation}
{-w}^{FE}_f\left({\ \sigma }_{f\ }\right)\mathrm{\ }=\frac{{\alpha }_{FE}}{2}{\sigma }^2_f+\ \frac{{\beta }_{FE}}{4}{\sigma }^4_f+\ \frac{{\gamma }_{FE}}{6}{\sigma }^6_f
\label{eq:UFEsigma}
\end{equation} 
(with ${\alpha }_{FE}<0$ and ${\beta }_{FE}$ or ${\gamma }_{FE} > 0 $ ) which does not represent a FE (or any other known) material. The functional form adopted for ${-w}^{FE}_f\left({\ \sigma }_{f\ }\right)\ $in these models corresponds actually to that of $u^{FE}_b{(P}_{FE})$ given in \eqref{eq:UFE}. Built into this (incorrect) assumption is already the possibility of NC, since the function described by Eq. \eqref{eq:UFEsigma} has regions of negative second derivative. Combining Eqs. \eqref{eq:UDEsigma} and \eqref{eq:UFEsigma}, we see that at small ${\sigma }_f$ the total system energy of the free charge configuration is:
\begin{equation}
-W^{DE}_f\left({\ \sigma }_{f\ }\right)-W^{FE}_f\left({\ \sigma }_{f\ }\right)\approx \left(\frac{d_{DE}}{2\epsilon }+\frac{{\alpha }_{FE}d_{FE}}{2}\right){\sigma }^2_f\ \ A
\label{eq:Utotsigma}
\end{equation} 
where $d_{DE}$ and $d_{FE}$ are the thickness of the DE and FE layers. The incorrect assumption for the functional form of $-W^{FE}_f\left({\ \sigma }_{f\ }\right)$ leads to the incorrect conclusion that, under the condition (termed ``capacitance matching''): ${d_{DE}}/{\epsilon}>{|{\alpha}_{FE}|d_{FE}}$, a stable configuration can be obtained at ${\sigma }_f=0$ with dielectric-like behavior for the combined system, but with a larger capacitance than that of the DE capacitor (concluding that the FE capacitor has NC). This conclusions cannot be obtained if the correct form of $-W^{FE}_f\left({\ \sigma }_{f\ }\right)$ (which always has positive second derivative, see Fig. \ref{fig:fig_4}) is used in the analysis. We conclude that the $Q^{FE}_f-V_{FE}$ behavior shown in Fig. \ref{fig:fig_1}a is unphysical for FE while Fig. \ref{fig:fig_1}b represents the expected behavior.

\begin{figure}[hb]
	\centering
		\includegraphics[width=0.3\textwidth]{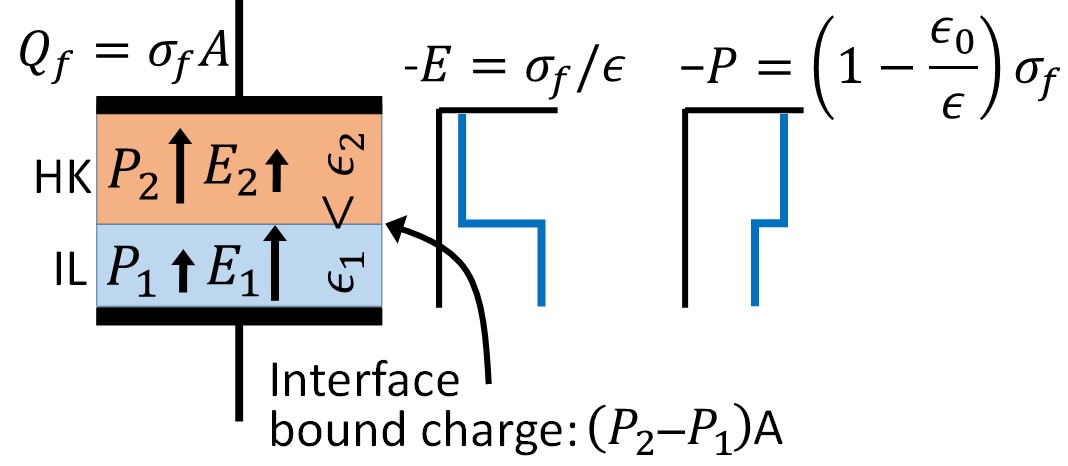}
	\caption{Schematic of interface layer SiO$_{2}$ (IL) and high-K linear dielectric (HK) bi-layer capacitor. The electric field is stronger in the IL, due to the polarization charge at the interface between the two layers, which results from the difference in polarization between the layers.}
	\label{fig:fig_6}
\end{figure}

We now consider bi-layer capacitors. At most interfaces, and particularly at incoherent/disordered interfaces between dissimilar materials, there are typically discontinuities in the polarization resulting in a net interfacial polarization charge (e.g. at the SiO$_{2}$/high-K dielectric interface in typical CMOS gate stacks (Fig. \ref{fig:fig_6}), the field is stronger in the SiO$_{2}$ layer due to the interface polarization charge). 

We now turn to the second case of interest: a bi-layer capacitor in which one layer is a DE (or even a semiconductor) and the other layer is a FE; systems for which there have also been numerous reports suggesting the stabilization of QSNC in the FE layers \cite{salahuddin_use_2008,krivokapic_14nm_2017,kwon_improved_2018,majumdar_revisiting_2016,dong_simple_2017}. Here, the total free energy of the material stack is typically assumed to be given by:
\begin{equation}
\begin{aligned}
& U^{system}_b=A \Bigg[\big(\frac{\alpha_{DE}d_{DE}}{2}P^2_{DE}\big) \\
& + d_{FE}\big(\frac{{\alpha }_{FE}}{2}P^2_{FE} +\frac{{\beta }_{FE}}{4}P^4_{FE}+ \frac{{\gamma }_{FE}}{6}P^6_{FE}\big)\Bigg]
\end{aligned}
\end{equation} 
The DE is in a stable configuration at small $P_{DE}$, but the FE has an unstable configuration (since ${\alpha }_{FE}<0)$ at small  $P_{FE}$. The ansatz, which we refer as ``QSNC ansatz 2'', is that the energies are added as function of a single polarization \cite{islam_khan_experimental_2011,appleby_experimental_2014}, which requires the polarization of the FE and DE layers to be the same, i.e. $P_{FE}=P_{DE}=P$, which is explicitly assumed in\cite{islam_khan_experimental_2011}. Under this assumption, the total system behaves like a dielectric (energy minimum at $P=0$), with a larger capacitance than a capacitor having only the DE layer, if the condition termed ``capacitance matching'' is met: ${\alpha}_{DE}d_{DE}>|{\alpha}_{FE}|d_{FE}$. This reasoning suggests the possibility of stabilized QSNC in MOS devices with a FE layer in the gate stack. Note, however, that even if the channel and the FE layer would have the same polarization, this would not lead to a sub-60 mV/dec SS as observed in devices with DE IL/HfO${}_{2}$-based FE bi-layers in the gate stack \cite{sharma_impact_2017,sharma_time-resolved_2018}; rather, it would lead to the channel and FE having similar dielectric-like behavior with the same permittivity (both having positive capacitance), which does not lead to sub-60 mV/dec SS. In any case, there is no general physical foundation for assuming the polarizations to be the same in all situations.
\noindent There are systems, however, in which a strong interfacial polarization coupling may lead to a uniform polarization across different layers, such as in multilayer epitaxial perovskites including FE layers \cite{ho_tsang_structure_2004,zhou_enhancement_2010}, which may be attributed to strong interaction of electrical and mechanical properties in these materials. This was precisely the type of system used in\cite{islam_khan_experimental_2011}. It is possible, however, for many interfaces (in particular incoherent-disordered interfaces), to exhibit no significant polarization coupling, with each layer adopting its optimal polarization independently (as in the example of SiO${}_{2}$/high-k dielectric stacks). 

Finally, we show that a QS ``apparent'' NC of the FE layer is possible in the case of a FE-DE bi-layer with strong interfacial polarization coupling. This analysis requires the FE layer to maintain it's FE phase and properties when in the stack (non-trivial assumption). In the simplest form (e.g. for thin layers), the total system energy in the absence of an external field is modeled as:
\begin{equation}
\begin{aligned}
& U^{system}_b= A \Bigg[\frac{\lambda }{2}{\left(P_{DE}-P_{FE}\right)}^2 + \Big(\frac{\alpha_{DE}d_{DE}}{2}P^2_{DE}\Big) \\
& +d_{FE}\Big(\frac{{\alpha }_{FE}}{2}P^2_{FE}+\ \frac{{\beta }_{FE}}{4}P^4_{FE} + \frac{{\gamma }_{FE}}{6}P^6_{FE}\Big)\Bigg]
\end{aligned}
\end{equation} 
\noindent where $\lambda $ ($>$0) is the interface polarization coupling parameter describing the strength of the coupling \cite{ho_tsang_structure_2004,zhou_enhancement_2010}. Polarization coupling in epitaxial perovskite superlattices may be attributed to electrostatic contributions, mechanical effects, strain, interface chemistry and interface structure \cite{dawber_unusual_2005,salev_polarization_2016}. This expression has a minimum at $P_{FE}=0=P_{DE}$, if the following conditions are met:
\begin{subequations}
\begin{align}
\alpha_{DE}d_{DE} &> \frac{|\alpha_{FE}|d_{FE}\lambda}{ \lambda - |\alpha_{FE}|d_{FE}} \label{eq:C1} \\
\lambda &>  |\alpha_{FE}|d_{FE} \label{eq:C2}
\end{align}
\end{subequations}
Under these conditions, both the FE and DE layer show minima at 0 polarization (DE-like behavior). Condition \eqref{eq:C1} reduces at large $\lambda$ to the ``capacitance matching'' condition of previous models. However, an additional, non-trivial requirement is needed: strong interfacial polarization coupling between the layers, expressed by condition \eqref{eq:C2}. Under these conditions, the FE layer does not have a NC, in fact, both the FE and DE layers have positive capacitance with similar permittivities. The very strong coupling forces the polarizations of both layers to be the same, so the system now behaves with just one polarization response, i.e. behaves electrically as a single material with a single permittivity. The system may exhibit a positive overall capacitance larger than that of a capacitor having only the DE layer, i.e. a ``stabilized'' QS ``apparent'' NC for the FE layer. This may be useful for applications in ultra-low EOT devices, but does not lead to NS QS sub-60 mV/dec SS. 

This clarifies the paradox of the ``stabilization'' of a microscopic atomic configuration at a point of unstable equilibrium: in thin FE layers adjacent to a DE with strong interfacial polarization coupling, the unstable configuration ($P_{FE}=0$) in the FE becomes stable due to the additional force fields due to the interfacial coupling to the polarization of the DE. 

In summary, we showed that  models of stabilized QSNC are either incorrect or not applicable to obtain NS QS sub-60 mV/dec SS in MOS devices. We proposed a model that sets the requirements for the observation of ``apparent'' QSNC of the FE layer in a DE-FE bi-layer stack which, in addition to the ``capacitance matching'' condition, requires strong interfacial polarization coupling between the FE and DE layers, and results in both the DE and FE layers behaving like dielectrics (positive capacitance for both layers) and may be useful to achieve ultra-low EOT devices.



%
%

%


\bibliography{ms}

\end{document}